**An Investigation into Outlier Elimination and Calculation Methods in the Determination of Reference Intervals using serum Immunoglobulin A as a model data collection**

Aidan Zellner[1], Alice M. Richardson[1], Brett A. Lidbury[1], Peter Hobson[2], Tony Badrick[3]

1. National Centre for Epidemiology & Population Health, Australian National University, Canberra ACT 2601, Australia

2. Sullivan Nicolaides Pathology, Brisbane QLD 4006, Australia

3. Biomedical Science, Bond University, Gold Coast QLD 4226, Australia and RCPA Quality Assurance Programs, Sydney NSW 2065, Australia

**Corresponding author**: Tony Badrick, RCPAQAP, Suite 201, 8 Herbert Street, St Leonards NSW 2065. Phone 1300 78 29 20. Email Tony.Badrick@rcpaqap.com.au

Short title: **Methods of Reference Interval Calculation with IgA**

**Abstract**

Background: Reference intervals are essential to interpret diagnostic tests, but their determination has become controversial.

Methods: In this paper parametric, non-parametric and robust reference intervals with Tukey and block elimination are calculated from a dataset of over 32,000 serum immunoglobulin A (IgA) measurements.

Results: The outlier elimination method was significantly more determinative of the reference intervals than the calculation method. The Tukey elimination procedure consistently eliminated significantly more values than the block method of Dixon and Reed across all age ranges. If Tukey elimination was applied, variation between reference intervals produced by the different calculation methods was minimal. Block elimination rarely eliminated values. The non-parametric reference intervals were more sensitive to outliers, which in the IgA context, led to higher and wider reference intervals for the older age groups. There were only minimal differences between robust and parametric reference intervals.

Conclusions: This suggests that Tukey elimination should be preferred over the block D/R method for datasets similar to the one used in this study. These are predominantly new observations, as previous literature has focused on the calculation technique and not discussed outlier elimination. This suggests the robust method is not advantageous over the parametric method and therefore due to its complexity is not particularly useful, contrary to CLSI Guidelines.

# Introduction

Reference intervals play a central role in the interpretation of diagnostic results. A patient's test result is often interpreted by comparison to a reference interval, two numbers specifying the range within which a 'healthy' individual's analyte value should lie. Hence the calculation of reliable reference intervals is of great importance. Traditionally reference intervals are either determined by a pathology laboratory itself or taken from literature[1] and are derived in a variety of ways.

In other words, despite the importance of reference intervals, no standardised method exists for their calculation. In fact, a wide variety of approaches are used. The Clinical and Laboratory Standards Institute (CLSI) is considered a peak body in the area and states reference intervals have traditionally been 'poorly' defined and not calculated by any 'uniform process'.[2] The CLSI has called for more research into whether these different statistical methods of outlier removal and calculation produce significantly different reference intervals for different analytes, or whether particular approaches are preferable in certain circumstances.[1]

The CLSI has described the most common methods of reference interval calculation in their recent Guidelines.[2] They state that reference intervals are directly calculated as the central 95% of measurements of a clinically screened healthy population. The Guidelines state determination is generally a three-step process: partitioning, outlier elimination and calculation. The first step, partitioning, involves separating the data into appropriate age and sex categories, as healthy levels often vary significantly across age and sex. The Guidelines have some minimal guidance on partitioning, but do not advocate any standardised method, instead noting that it is a highly contextual process and specific to the structure of the dataset. The second step, outlier elimination, is required, as clinical screening is normally insufficient to eliminate all unhealthy individuals from a dataset, particularly those with subclinical illness. While small in number, these individuals can have a significant effect on the reference interval, hence their systematic removal is required for the production of valid reference intervals. The two methods advocated by the CLSI[2] are the block method of Dixon and Reed (block D/R)[3,4] and the Tukey elimination procedure.[5] The third step is the reference interval calculation method itself. Three general approaches are used: parametric[6], non-parametric[5] and robust.[7]

The view that the only acceptable method to determine a reference interval is the direct approach, wherein healthy individuals are recruited, re-assessed for health, their samples analysed and the results described using parametric or non-parametric statistics, is changing. Indirect methods seek to distinguish reference populations within the mixture of populations that exist in a laboratory database.[8,9] Sikaris[10] argues that the indirect method is superior to the direct method because 'that reference populations defined within a laboratory database are better candidates for comparison: they have had their samples collected, transported and analysed exactly as patients do, and they have similar complaints (e.g. abdominal pain) without the disease (e.g. pancreatitis)'.

Reference interval data collected on 11 commonly measured biochemical analytes using the indirect method[11] found that the results were not statistically or clinically different from the best available published population-based direct sampling studies when those results originate from the tests that are not typically ordered on "healthy" patients. Indeed, it appears that the main differences between directly and indirectly defined reference intervals are due to assay bias.[12]

In this work we have used the indirect method on a very large data base to provide a significant sample size of subject results to compare different methods of determining reference intervals. Other approaches to investigate the effects of outliers and calculation method have used simulation,[1] however using subject results has the advantage that the data are from a real population.

Comparison of the various approaches to outlier elimination and calculation method has been undertaken several times with varying lenses. Different markers have been studied ranging from highly specific markers such as high-sensitivity cardiac troponin[13] to standard full blood count analytes such as hemoglobin and hematocrit.[14] The amount of skewness and therefore transformation to Normality is part of some studies.[15,16] So also is outlier removal[14,16,17] but there is less literature on a comparison between methods of outlier removal. Different combinations of estimation method have been compared.[1,18]

For diagnostic purposes, elevated serum immunoglobulin (Ig) levels indicate an immune response, generally because of infection or an immune disorder. Immunoglobulin (Ig) serum profiles indicate the nature of the immune response experienced by the patient; for example, IgM elevation indicates an initial infection by a pathogen. Specific IgG responses indicate immune memory, and thus a secondary response to infection, and with IgM is diagnostically useful in terms of understanding whether reinfection has occurred. IgA is associated with mucosal infection. Depleted Ig levels are indicative of impaired immune function which occurs from a variety of disorders[19]. While Ig reference intervals are widely available online, there is scant description on their derivation. Some more general statistical analysis has been conducted on the relationship between Ig levels, age and sex.[20] On the other hand, methodological analysis work on reference intervals has tended to focus on other analytes, such as cardiac troponin I.[21] This paper will focus on IgA, as not only it is of great clinical significance in terms of its common use and its ability to indicate disease in humans, but it is also representative of many biomarkers which follow a non-Normal distribution with a long right tail.

**Methods**

*Data*: Reference intervals were calculated from a population of 37,074 pathology requests for community-living individuals in the states of New South Wales and Queensland in Australia, in 2015 - 2016. The dataset contained 32,293 serum IgA measurements, along with the individual's sex and age. Five individuals aged ≥ 100 years and three individuals with missing ages were not included in the analysis leaving 32,285 records for analysis. Clinical notes were available for 16,659 (49.4%) of the individuals.

*Age-sex partitioning*: The CLSI Guidelines recommend the Harris and Boyd test[2] to establish whether datasets should be analysed separately e.g. by age. The test was applied to the sexes separately, testing uniform age partitions with widths of 1, 2, 3, 5, 8, 9, 10, 15, 20, 30, 40 and 50 years. The test was also applied to a non-uniform partition with 1 year age segments from 1-15 and then 20 year segments after 15 years.

*Outlier elimination*: The two methods of outlier elimination commonly used in the literature are the block D/R[3,4] and Tukey elimination procedures.[5] The block D/R procedure eliminates values based on the D/R ratio. The D/R procedure eliminates a value $x$ when $D(x)/R > (1/3)$ where $D(x)$ is the difference between $x$ and the closest less extreme data value to $x$ and $R$ is the range of the dataset. When $x$ is less than the mean, $D(x)$ will be the difference between the $x$ and the next largest value in the dataset, when $x$ is greater than the mean, it will be the difference between $x$ and the next smallest value in the dataset. When a single value is eliminated, the 'block' of all more extreme values are also eliminated.

The Tukey method eliminates values $x$ when $x < Q1 – 1.5 \times IQR$ (*Inter Quartile Range*) or $x > Q3 + 1.5 \times IQR$, where $Q_1$ and $Q_3$ are the first and third quartiles and *IQR* is the interquartile range ($Q3$

– $Q1$).[4] Thus, the Tukey method eliminates all values more than 1.5 IQRs away from the first and third quartiles. The Tukey method assumes a degree of normality including symmetry hence before application, the data was tested for normality via the Anderson-Darling, Kolmogorov-Smirnov or skew normality tests recommended by the Guidelines.[2] If the dataset segment failed any of the tests with 95% certainty, the data was transformed using a Box-Cox transformation as per the Guidelines.[2] Box-Cox transformations are a family of curves[22] used to normalise data of the form $T(y) = \frac{(y+\lambda_2)^{\lambda_1}-1}{\lambda_1}$ determined by one or two parameters (the one parameter version has $\lambda_2 = 0$). Here one parameter transformations were used except for groups of observations containing IgA values of zero which required two parameters.

*Calculation method:* Three approaches to reference interval calculation are used in literature, parametric[6], non-parametric[5] and robust[7] methods.

At a conceptual level, the parametric method involves fitting a distribution to the dataset and then using the distribution to calculate the reference interval. The parametric method calculates the reference interval as the standard 95% confidence interval around the mean for some distribution. A Gaussian distribution is typically used and recommended by the Guidelines, but other distributions or transformation of the data are required for non-normal data. As the IgA data was not Normally distributed, transformation was used and Box-Cox transformations are considered one of the most flexible transformation families.

The non-parametric reference interval relies exclusively on the data presupposing no distribution. It takes the values 2.5% and 97.5% along the ranked observations according to magnitude. If these positions are non-integer, linear interpolation is used.

The robust method is a more complex procedure involving bi-weight estimator functions designed to be a compromise method with the advantages of both the parametric and non-parametric procedures. This method calculates the reference interval using a combination of bi-weight estimator functions weighted with a Students-t distribution and a tuning constant *c* which determines the vulnerability of the reference intervals to outliers. The robust method does not assume that the data follow a Normal distribution but does assume a symmetric distribution. Many of the IgA age groups are significantly positively skewed, hence Box-Cox transformation was employed before applying the robust method of calculation.

Firstly, a robust estimate of the centre of the values is calculated by the following algorithm

Set $T_{bi}$ = median of the observations

Calculate standardized residuals $u_i = \frac{x_i - T_{bi}}{c \times MAD\ /\ 0.6745}$

Calculate weights $w_i = \begin{cases} (1 - u_i^2)^2 & -1 < u_i < 1 \\ 0 & otherwise \end{cases}$

Update $T_{bi} = \frac{\sum w_i \times x_i}{\sum w_i}$

Stop when the change in consecutive values of $T_{bi}$ is negligible (< 0.001% is recommended [2]).

A typical value of c is 3.7.[2] This value has been found to work well in simulation studies.[7] Note also that only 0.0001 of a Normal distribution lies beyond 3.7 standard deviations from the mean. The bi-weight function that produces the weights $w_i$ is a redescending curve (see Figure 1) that applies a weight to each observation before proceeding to calculate the robust measure of centre. Standardised residuals that are close to zero are kept unchanged, which applies to the bulk of the data. Mild outliers are reduced in size, thereby reducing their contribution to the calculation. Strong outliers are set to zero, effectively eliminating their effect on the calculation.

Once $T_{bi}$ has been obtained, a measure of standard error is obtained to create the upper and lower limits of the robust reference interval. The standard error is in two parts, mirroring the formula for the prediction interval for a single observation in simple linear regression[5] is the sum of two measures of variation referred to as $s_{bi}$ and $s_T$. The first, $s_{bi}$, is a robust measure of spread whose formula involves the median and median absolute deviation; while the second, $s_T$, is a robust measure of spread involving $T_{bi}$ and $s_{bi}$.[23] A second tuning constant is required for the calculation of $s_{bi}$, which is set at 205.6 [2] which is the value needed to achieve asymptotic efficiency of 95%.[7]

In summary, for a sample $X = x_1, ..., x_n$, the reference intervals are defined as below.[2]

Parametric: $\bar{x} \pm 1.96 \times \sigma_X$

Non-parametric: $x_{0.025(n+1)}$ and $x_{0.975(n+1)}$ with linear interpolation

Robust: $T_{bi} \pm StuT(0.025, n-1)(s_{bi}^2 + S_T^2)^{1/2}$

where $T_{bi}$, $s_{bi}$, and $S_T$ are defined above and StuT(0.025, n-1) is the upper 2.5[th] percentile of a Student t distribution with n – 1 degrees of freedom where n is the sample size.

R statistical software[24] was used to determine nine reference intervals for each age group: all possible combinations of outlier elimination method (block D/R, Tukey, none) and calculation procedure (parametric, non-parametric, robust) discussed in the Guidelines (i.e. block D/R parametric, block D/R non-parametric, block D/R robust, Tukey parametric, Tukey non-parametric, Tukey robust, none parametric, none non-parametric, none robust).[2] The parameters for the Box-Cox transformation were estimated with a log likelihood plot using `boxcoxfit` in the R library `geoR`.[25] The relevant outlier elimination methods were applied to the transformed data and the data was then back-transformed with the outliers removed.

**Results**

A descriptive summary of the clinical conditions of the subjects will be present first, followed by the need for Box-Cox transformation and the results of the age-sex partitioning. The section concludes with the reference intervals for IgA in the chosen age-sex groups.

*Clinical conditions*

Clinical notes were available for 16,659 (49.4%) of the individuals, and 7,974 (24.7%) of the individuals had clinical notes which could be classified. Of the classified clinical notes, 2,073 (26.0%) referred to multiple myeloma and monoclonal gammopathy of undetermined significance; and 1,874 (23.5%) referred to lymphoproliferative disorder. Whilst there is no evidence that these percentages would be maintained in the population where notes could not be classified, the nature of the most prevalent conditions mentioned suggests that the individuals in this database represent a population that are as close to a clinically screened healthy population as could be expected.

*Need for Box-Cox transformation*

Figure 2 provides a visual characterization of the IgA results for one-year age group from 0 to 100 years. Most of the distributions by year are skewed (Anderson-Darling test $p < 0.05$ for 87 female age groups with another 4 missing due to small sample size; and $p < 0.05$ for 78 male age groups with another 4 missing due to small sample size). Figure 3 shows the skewness in the ten-year age groups which will be used in the subsequent calculation of reference intervals (see below). Table 1 shows the summary statistics for the same ten-year age groups.

*Age-Sex Partitioning*

*Age Partitioning:* The five individuals with age ≥ 100 years were excluded due to the small number of individuals. For partitioning into 1 year age intervals, the Harris and Boyd test indicated that most of the ages of 1 to 20 years should be partitioned (threshold reached). For ages ≥ 20 years, the results for all partitions tested (see Methods) indicated no need for partitioning (threshold not reached). A compromise between a partition sufficiently fine to show the differences in the early ages, but not too fine a partitioning of the middle and older age ranges, will best reveal the patterns in IgA levels across age groups. Therefore a ten-year partition of age with 10 year groups of the form $[10i, (i + 1)10]$ for $i = 0, \ldots, 9$ has been chosen for this paper.

The need for some age partitioning is supported by the clear relationship between IgA levels and age visible in Figure 1. As age increases the mean and spread of the IgA values increases significantly. There is a rapid increase in mean IgA concentration at lower ages from 1 to 20 years, which levels out for > 20 years. These results also accord with previous findings.[9] For the ≥ 90 years age group, IgA seems to increase. However, the dataset had fewer individuals in this age range, and the confidence intervals in Figure 4 overlap, so the trend is not statistically significant.

*Sex Partitioning:* Sex partitioning was not used to calculate the reference intervals, as it was only recommended by the Harris-Boyd test for between 2% and 10% of all the different age partitions tested. This accords with the overlapping boxplots by sex observable in Figure 2.

*Reference Intervals*

Table 2 contains the calculated reference intervals for IgA. The midpoint of IgA reference intervals are consistently higher in women, particularly for ≥ 60 years age groups. However, this change is small; on average approximately 10% of the IQR across the age segments.

The reference intervals rise and expand with age to roughly double their initial width. Differences between calculation method are minimal for the early ages. However, as age increases the reference intervals calculated via different methods separate.

Figures 4 - 5 shows the midpoints and widths of all the calculated IgA reference intervals with symbols representing the different outlier elimination methods. The graph is grouped substantially by outlier elimination method as opposed to by calculation method. This demonstrates that the elimination method is the dominant determinant of the reference interval. Once a given outlier elimination method is used, the various calculation methods only result in small differences to the reference interval. When the same calculation method is used with two different outlier elimination methods, there is still often a significant difference in the reference interval.

## Discussion

The determination of reference intervals involves two different selection processes which feed into each other. One process, a clinical process, involves the selection of a reference population with reasonably well articulated health/disease status. But in reality, it is difficult to define this status because of subclinical illness or unexpected results in a particular test in a particular subject. These results are then analysed statistically to produce a reference interval, usually on the assumption by the statistician that they come from a homogeneous population. So, while the clinicians may produce a less than perfect sample hoping that the statistician will remove any outliers, and the statistician may accept the sample assuming that it has been selected so that it is homogeneous. The emphasis has been therefore on the statistical method to determine the limits of the reference interval rather than on the method to remove potential outliers.

Yet the outlier elimination method is more determinative of the reference interval than the calculation method (Figures 4 - 5). This is a new observation not reported in literature. Previous studies have primarily examined the calculation method, and simply used either the block D/R method or Tukey method, not compared the two and not reported the number or proportion of outliers eliminated.[1]

The Tukey method is a significantly more aggressive outlier eliminator than the block D/R procedure. Table 1 shows that the block D/R method eliminates only one individual in the 20 – 30 years age group. For the remaining age groups, it eliminates zero individuals. Hence it is nearly equivalent to the "no elimination" data. The Tukey method removes between 2% and 5% of the data for each age group. The elimination of the Tukey method is so substantial that there is little observable difference between the reference intervals produced by the various calculation methods, once Tukey elimination is used. The maximum difference between lower and upper reference interval values is 2% across all different IgA Tukey reference intervals. Differences of up to 6% occur between the D/R elimination reference intervals. The block D/R method has been described as highly conservative, even by the Guidelines,[2] hence this is not surprising, however no previous study seems to have observed such an extreme lack of elimination.[15] Hence when using D/R elimination care should be taken to ensure it is eliminating observations that can truly be regarded as outliers.

This convergence of calculation method implies the Tukey method is superior for the current dataset, though as in all datasets it is not possible to tell if the outliers removed are all genuine outliers. The various calculation methods represent significantly different mathematical processes. The fact that these converge to a very similar reference interval is highly indicative that a fundamental characteristic of the dataset is being discovered. It is possible that the Tukey method is being overly aggressive for this analyte. There is the possibility of tuning the Tukey and D/R methods by varying the widths of elimination by either changing the cutoff from the default of 1/3 for the D/R method or changing coefficient of 1.5$IQR$ range for the Tukey method.[2] However, these variations do not seem to be in common practice in the literature, hence if choosing between the default block D/R and Tukey elimination methods, the Tukey seems preferable.

The calculation method does have some effect on the reference intervals, despite the dominance of the outlier elimination method. The non-parametric reference intervals are consistently more susceptible to outliers. In the IgA context where most of the age groups are skewed by positive outliers, this means the reference intervals are wider and higher. In Figures 4 - 5, as the spread of the IgA values increases with age, the upper bound of the non-parametric reference interval increases significantly above the upper bounds of the parametric and robust reference intervals. The results make sense in terms of the non-parametric algorithm employed. The procedure takes the values that are positioned at the 2.5$^{th}$ and 97.5$^{th}$ percentiles whereas the robust and parametric approaches involve fitting a symmetric distribution to the data. Thus, it is not surprising that the non-parametric method, which makes no assumptions about the underlying data structure, is more susceptible to large proportions of outliers.

The robust and parametric methods produce very similar reference intervals. Across the reference intervals in Table 1, the robust and parametric values do not differ by more than 2%. This is not surprising as both methods include a Box-Cox transformation, meaning that the input data to both the robust and parametric methods is very similar. This is despite both the parametric and robust reference intervals both possessing a variety of distribution choices (parametric) and tuning constants (robust) leading to the reference intervals being able to be calculated in a variety of ways. We have based the parametric approach on a Normal distribution for IgA (following Box-Cox transformation where necessary), but other distributions can be used e.g. Gamma, Beta. Also, with the robust method, there are two tuning parameters to select. Some studies recommend estimating the tuning parameters with likelihood ratio tests,[15] which was not done here. Calculation of the robust reference interval already required significant iteration to calculate the $T_{bi}$, $s_{bi}$ and $S_T$ values, and two likelihood ratio tests to find the optimal parameters for the Box-Cox transformation. Adding additional parameter fitting would increase the complexity of calculating the robust reference interval vastly beyond that of the other two methods. Hence for the large IgA dataset in this study, the robust method does not seem to have any advantages over the parametric methods.

This departs from the recommendations of the Guidelines which are enthusiastic about the robust method as a compromise method with the best aspects of both the parametric and non-parametric approaches. The Guidelines' recommendation is perhaps motivated by more practical considerations around the use of small datasets by pathology labs, for which the robust method may still have some advantages. However, the recommendation is unrealistic in the sense that the robust method is not implemented in standard spreadsheet software. The attractiveness of easy-to-calculate methods has been noted[13] and others[26] have advocated for the use of statistical software to allow for more complex calculations to be widely used.

### Conclusions

This paper has used IgA as an illustrative example because the distribution of values observed in this large dataset pose special problems in outlier elimination and reference interval calculation. Other immunoglobulins will be addressed in future research, and the methods of this paper could be applied to other analytes with differing degrees of skewness e.g. ALT, CRP. The strength of this analysis is the sample sizes across all age/gender cohorts.

This population can be characterized as an indirect sample because it is a community cohort who may be healthy or may be living with pre-clinical, subclinical or clinical conditions that lead to elevated IgA. Nonetheless this population is exactly that from which a typical subject presenting a sample to a community laboratory belongs. The production of reference intervals for this population is therefore of value, and statistical methods for direct samples can be translated to indirect samples [Jones et al].

There are possibly some statistically significant differences with regard to sex and IgA levels which merit further research, and perhaps for the production of high quality reference intervals for clinical practice, sex partitioning may be preferable.

An interesting area for further research would also be to calculate robust and parametric reference intervals with newer outlier detection methods[28] and observe whether there is significant variation between them.

The findings of this paper indicate that the complex calculations involved in the robust reference intervals may not be necessary as long as stringent outlier elimination is conducted. The reference intervals for IgA are substantially the same whichever calculation method is used, once an outlier elimination method is established. Results for other Ig types will be explored in a future paper.

**Figures**

**Figure 1**. The Tukey biweight function.

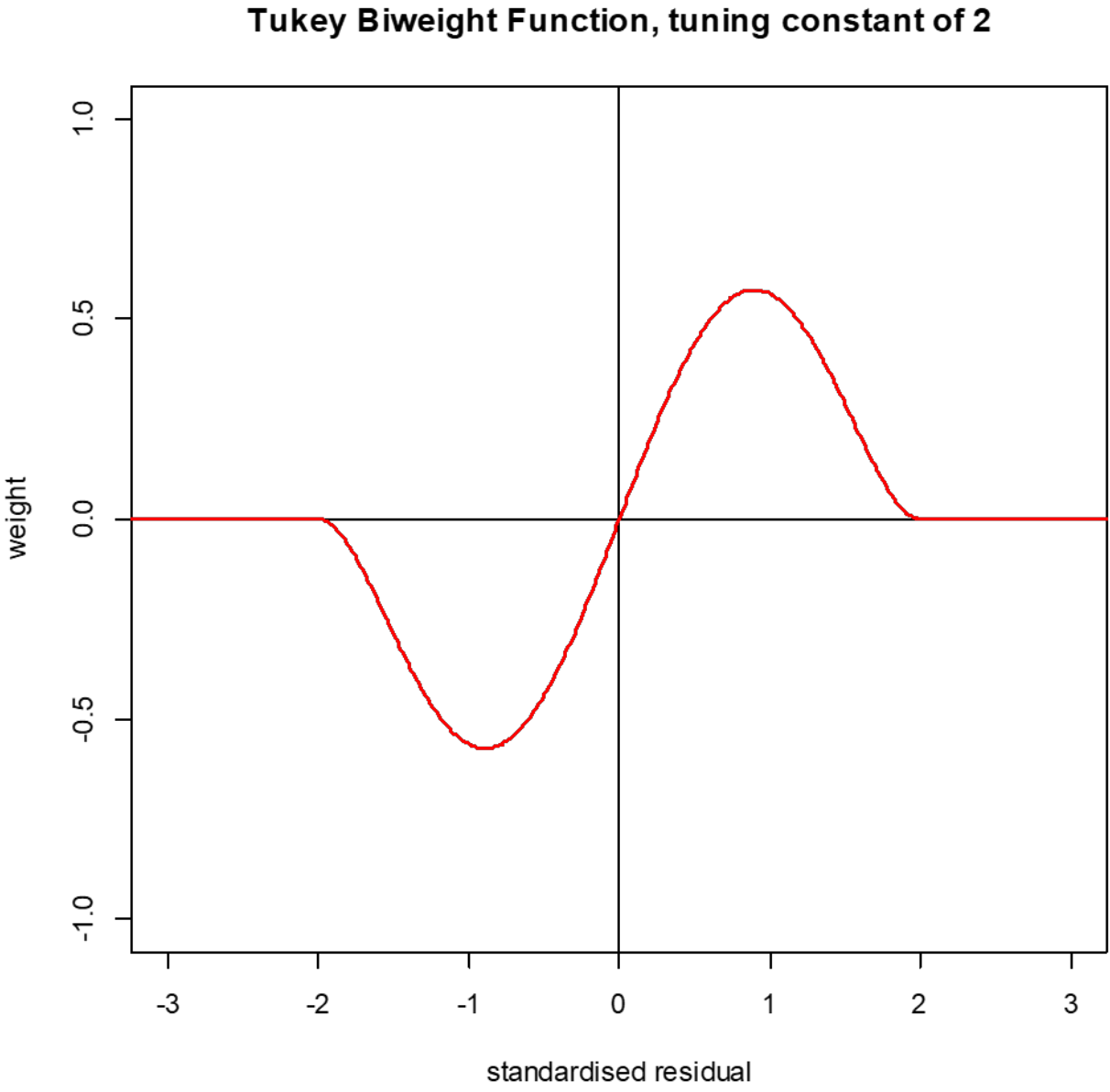

**Figure 2:** Box plots of the values of IgA used to calculate the reference intervals, plotted in one year age groups. Red box plots are used for the female datasets and black for the male datasets. The central thick lines represent the dataset mean. The thinner box lines represent the datasets quartiles and the whiskers 1.5IQR above and below the quartiles. The y-axis has been truncated at 6 g/L in the lower plot. Individuals (n = 5) aged over 100 years have also been excluded. (IgA = Immunoglobulin A, g/L = grams per litre).

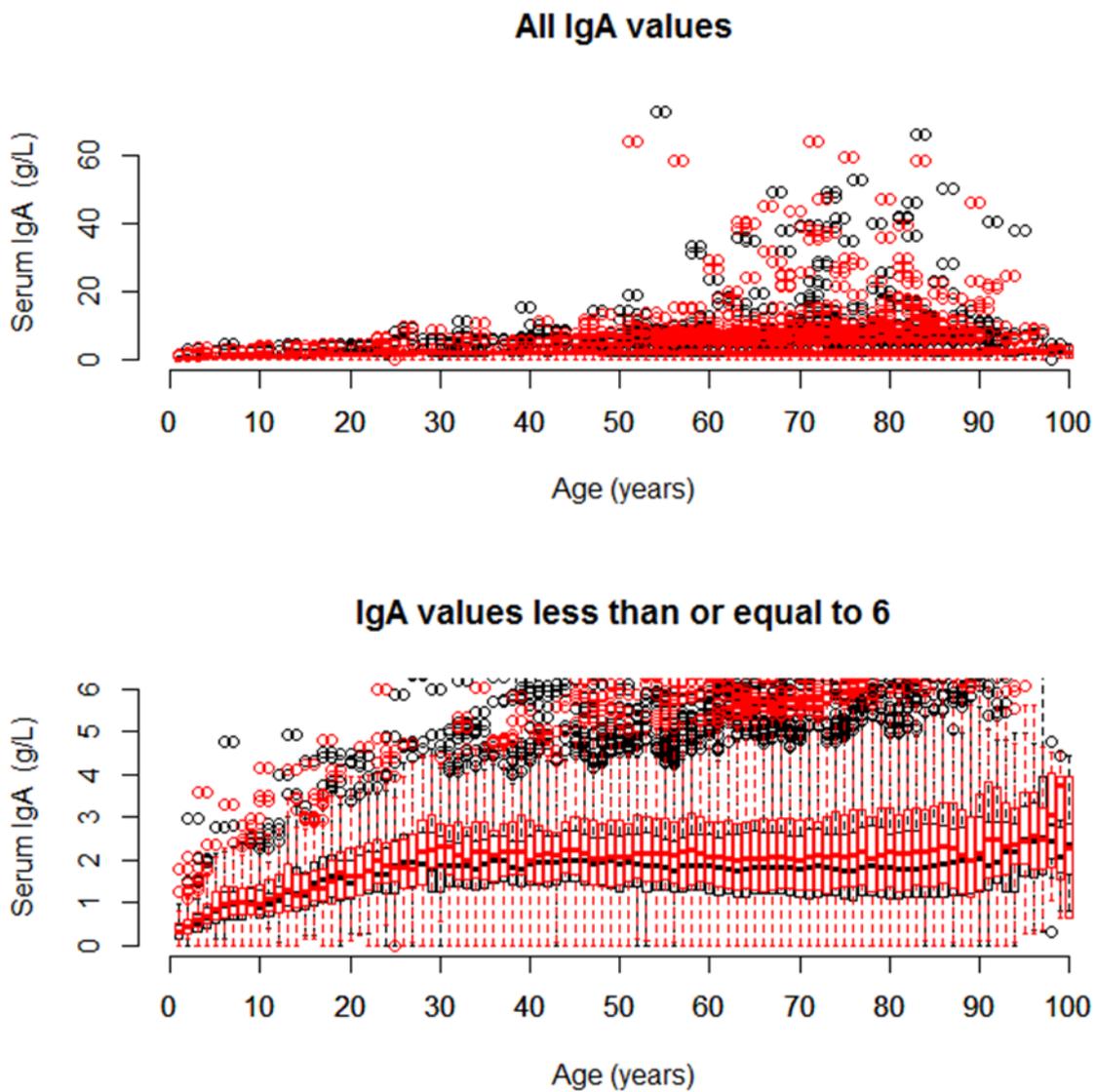

**Figure 3:** Normal Q-Q plot of IgA levels in ten-year age groups. g/L = grams per litre.

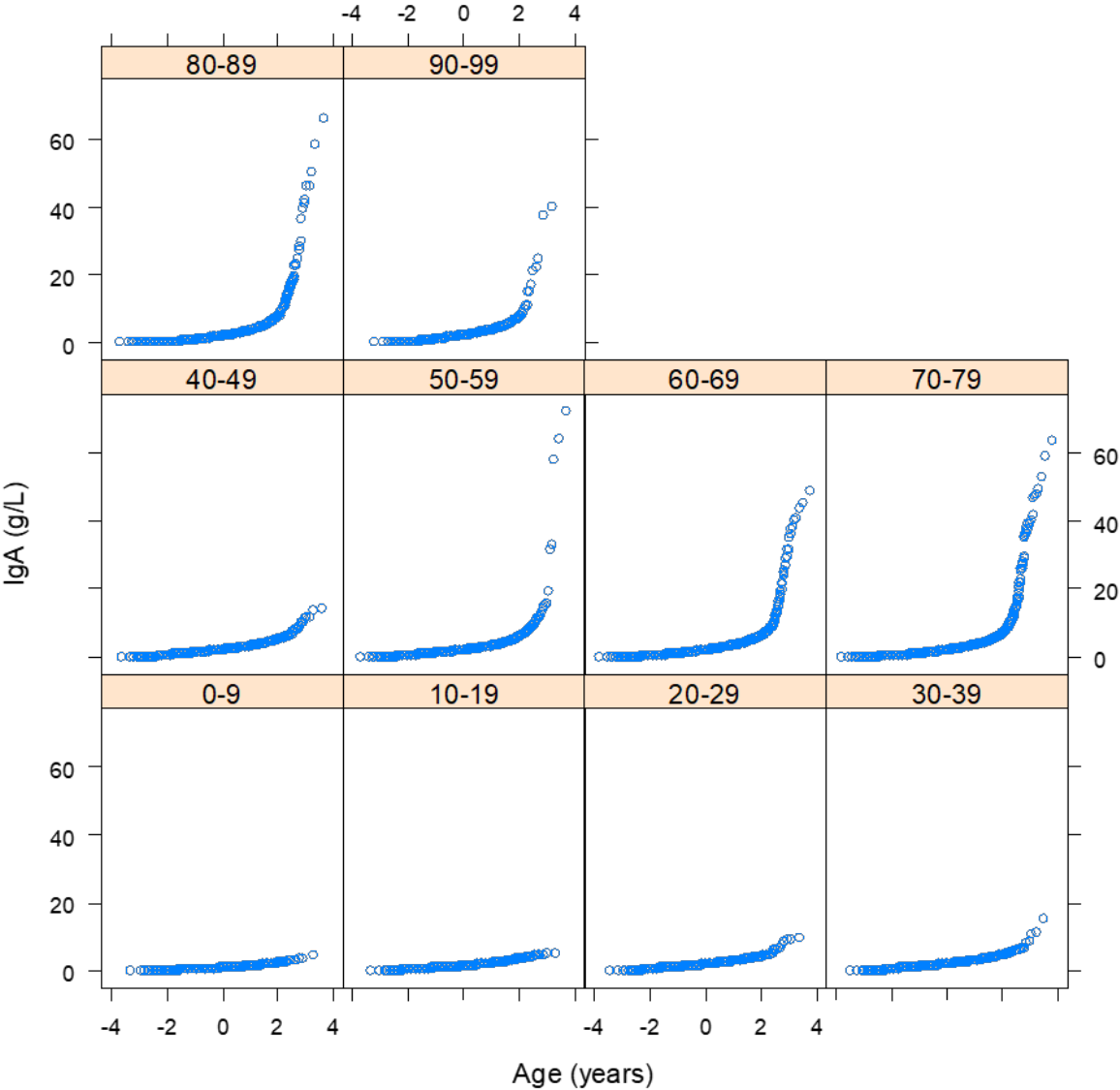

**Figure 4:** IgA Reference Intervals shown alongside their respective populations in grey plotted against age with a 10 year wide age partition. IgA values are truncated at 10 g/L. Each colour represents the different calculation methods (e.g. red lines on the second graph are non-parametric reference intervals with Tukey elimination). The reference intervals expand and rise as age increases. The differences between calculation methods also increase with age as the number of extreme measurements increase (IgA = Immunoglobulin A, g/L = grams per litre).

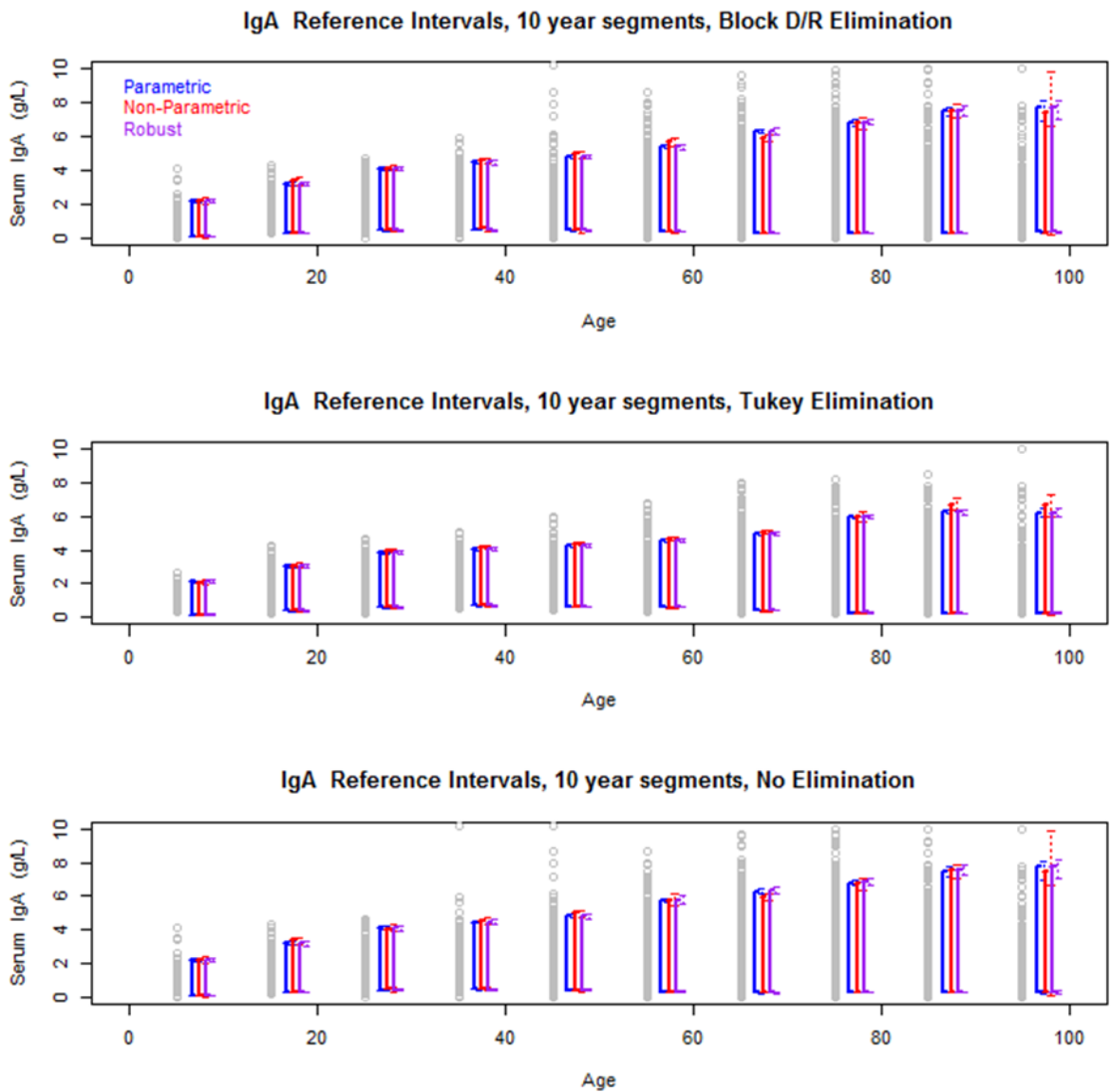

**Figure 5:** Midpoints and widths of the IgA reference intervals measured in grams per litre. Each symbol represents a different outlier elimination method and each colour a different calculation method (e.g. red triangle is the non-parametric Tukey reference interval). Coloured lines connect the same outlier elimination and calculation methods for visual clarity. The graph is dominantly grouped by shape not colour. This demonstrates that the elimination method is the dominant determiner of the reference interval (IgA = Immunoglobulin A, g/L = grams per litre).

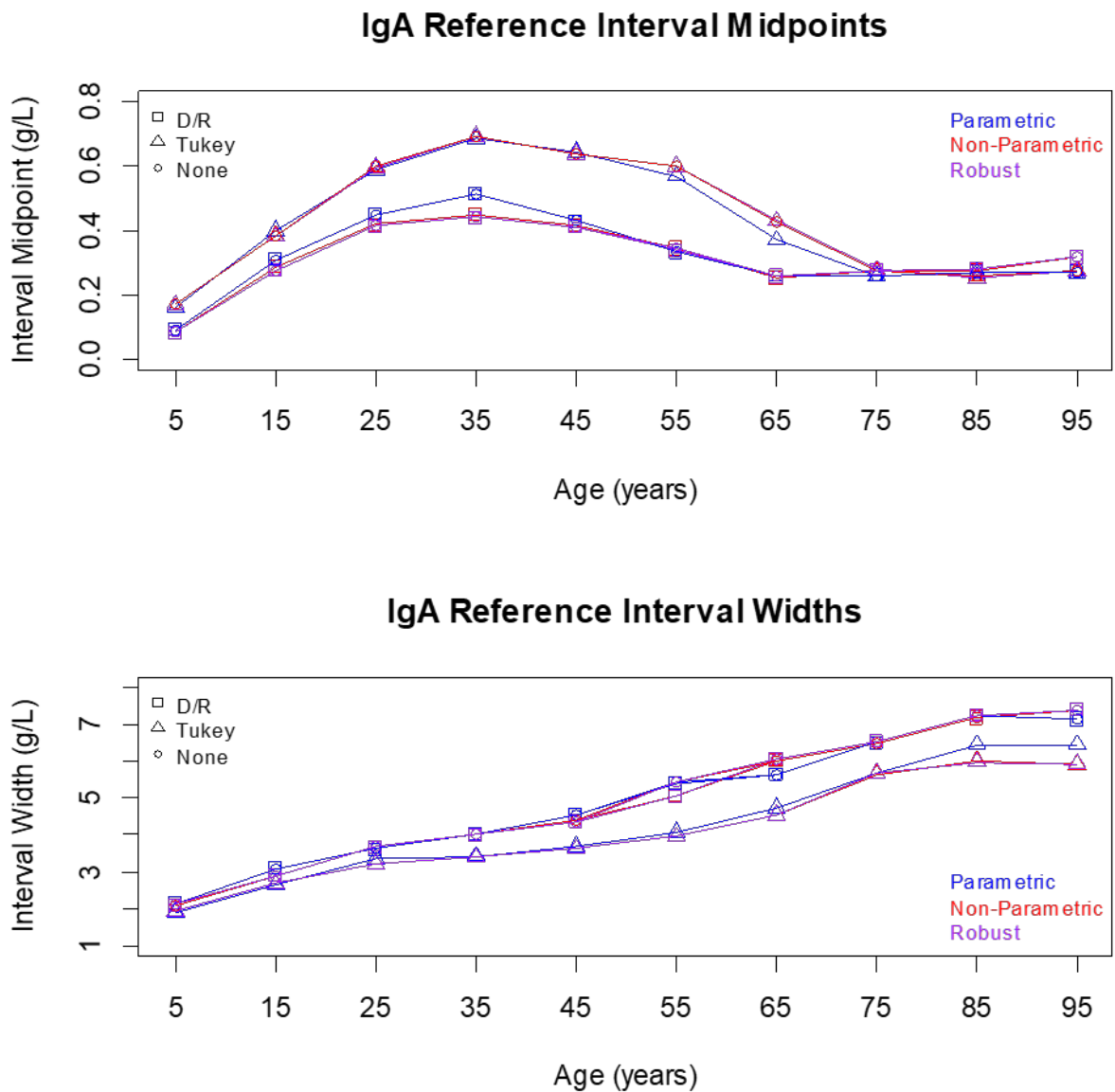

| Age group (years) | N (F) | Mean (SD) | N(M) | Mean(SD) |
|---|---|---|---|---|
| 0 – 10 | 371 | 0.86 (0.59) | 516 | 0.87 (0.53) |
| 10 – 20 | 638 | 1.47 (0.76) | 520 | 1.44 (0.77) |
| 20 – 30 | 1061 | 1.90 (0.92) | 512 | 2.08 (1.09) |
| 30 – 40 | 1453 | 2.03 (1.07) | 722 | 2.30 (1.31) |
| 40 – 50 | 1895 | 2.12 (1.11) | 1190 | 2.29 (2.49) |
| 50 – 60 | 2853 | 2.12 (2.03) | 1847 | 2.46 (2.49) |
| 60 – 70 | 3598 | 2.12 (2.00) | 3189 | 2.43 (2.56) |
| 70 – 80 | 3530 | 2.27 (2.82) | 3599 | 2.53 (3.00) |
| 80 – 90 | 2137 | 2.33 (3.06) | 1907 | 2.71 (3.07) |
| 90 – 100 | 441 | 2.52 (2.97) | 306 | 2.971(2.75) |

**Table 1:** Summary statistics for IgA values measured in grams per litre. N = number of individuals in the age group; F = female; M = male; SD = standard deviation.

| Age group (years) | N | D/R | D/R Para | D/R Non-para | D/R Robust | Tukey | Tukey Para | Tukey Non-Para | Tukey Robust |
|---|---|---|---|---|---|---|---|---|---|
| 0 - 10 | 887 | 0 | 0.083 – 2.179 | 0.092 – 2.226 | 0.083 – 2.184 | 26 | 0.170 – 2.100 | 0.160 – 2.043 | 0.170 – 2.107 |
| 10 - 20 | 1158 | 0 | 0.286 – 3.188 | 0.310 – 3.400 | 0.276 – 3.169 | 23 | 0.385 – 3.066 | 0.400 – 3.052 | 0.380 – 3.068 |
| 20 - 30 | 1573 | 1 | 0.420 – 4.104 | 0.450 – 4.069 | 0.415 – 4.095 | 34 | 0.599 – 3.818 | 0.590 – 3.950 | 0.596 – 3.816 |
| 30 - 40 | 2175 | 0 | 0.449 – 4.472 | 0.514 – 4.550 | 0.441 – 4.455 | 72 | 0.693 – 4.081 | 0.686 – 4.110 | 0.692 – 4.085 |
| 40 - 50 | 3085 | 0 | 0.418 – 4.805 | 0.430 – 4.989 | 0.411 – 4.785 | 107 | 0.641 – 4.269 | 0.645 – 4.345 | 0.639 – 4.269 |
| 50 - 60 | 4700 | 3 | 0.345 – 5.426 | 0.334 – 5.746 | 0.338 – 5.388 | 209 | 0.600 – 4.554 | 0.570 – 4.650 | 0.601 – 4.556 |
| 60 - 70 | 6787 | 0 | 0.253 – 6.286 | 0.260 – 5.903 | 0.258 – 6.331 | 236 | 0.429 – 4.967 | 0.370 – 5.110 | 0.430 – 4.975 |
| 70 - 80 | 7129 | 0 | 0.276 – 6.772 | 0.260 – 6.780 | 0.276 – 6.828 | 55 | 0.272 – 5.928 | 0.260 – 5.950 | 0.278 – 5.973 |
| 80 - 90 | 4044 | 0 | 0.276 – 7.472 | 0.270 – 7.529 | 0.280 – 7.547 | 32 | 0.256 – 6.267 | 0.270 – 6.730 | 0.252 – 6.238 |
| 90 - 100 | 747 | 0 | 0.318 – 7.717 | 0.270 – 7.422 | 0.318 – 7.717 | 8 | 0.275 – 6.201 | 0.270 – 6.725 | 0.273 – 6.212 |

**Table 2:** IgA Reference Intervals measured in grams per litre. N = number of individuals in the age group. The columns headed 'D/R' and 'Tukey' contain the number of individuals eliminated by each respective outlier elimination procedure (Para = Parametric, Non-para = Non-Parametric, Robust = Robust).